\documentclass[amsmath,amssymb,a4paper,showpacs,twocolumn,nofootinbib]{revtex4}
\usepackage{bm}
\usepackage{color}
\usepackage[pdfpagemode=None,pdfstartview=FitH]{hyperref}
\usepackage{mathbbol}
\newcommand{\be}{\begin{eqnarray} \begin{aligned}}
\newcommand{\ee}{\end{aligned} \end{eqnarray} }
\newcommand{\bc}{\begin{center}}
\newcommand{\ec}{\end{center}}
\newcommand{\half}{\frac{1}{2}}
\newcommand{\ran}{\rangle}
\newcommand{\lan}{\langle}
\newcommand{\im}{\text{\large $\mathbb{1}$}}
\newcommand{\Cn}{\mathbb{C}}
\newcommand{\re}{\mathop{\mathbb{R}}\nolimits}
\newcommand{\esp}{\mathop{\mathbb{E}}\nolimits}
\newcommand{\rlprt}{\mathop{\mathrm{Re}} \nolimits}
\newcommand{\imprt}{\mathop{\mathrm{Im}}\nolimits}
\newcommand{\tr}{\mathop{\mathrm{tr}}\nolimits}
\newcommand{\eval}[2]{\left. #1 \right \arrowvert _{#2}}

\newcommand{\iu}{\mathrm{i}}

\newcommand{\wavyline}[3]{
\multiput(#1,#2)(4,0){#3}{
\qbezier(0,0)(1,1)(2,0)
\qbezier(2,0)(3,-1)(4,0)}}
\begin{document}
\bibliographystyle{h-physrev}
\title{Estimation of unitary quantum operations}
\author{Manuel A. Ballester}
\email{ballester@math.uu.nl }
\homepage{http://www.math.uu.nl/people/balleste/}
\affiliation{Department of Mathematics, University of Utrecht, Box
80010, 3508 TA Utrecht, The Netherlands}
\pacs{03.67.-a}
\begin{abstract}
The problem of optimally estimating an unknown unitary quantum operation with the aid
of entanglement is addressed. The idea is to prepare an entangled pair, apply the unknown unitary to
one of the two parts, and then measure the joint output state. This measurement could be an entangled
one or it could be separable (e.g., measurements which can be implemented with local operations and classical
comunication or LOCC). A comparison is made between these
possibilities and it is shown that by using
nonseparable measurements one can improve the accuracy of the estimation by a factor of $2(d+1)/d$
where $d$ is the dimension of the Hilbert space on which $U$ acts.
\end{abstract}
\maketitle
\section{Introduction}
Consider a one-qubit unitary gate, the following
question arises: ``how to characterize it?'' This question is motivated
by recent experiments in quantum optics \cite{demartini:sqd}.
A possible approach is to prepare many known states and use them as inputs, and then
measure the outputs that they produce; this is known as
\emph{quantum process tomography} \cite{Nielsen:book}.
It turns
out that one needs as inputs a basis of the Hilbert space plus some
linear combinations thereof. The disadvantage of this
approach is that, in many practical situations,
such a set of states is not feasible in the
laboratory \cite{demartini:sqd}.

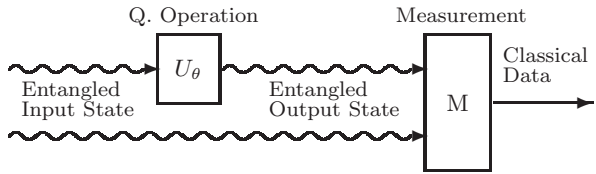
\begin{figure}[ht]
\begin{center}
\setlength{\unitlength}{.035in}
\begin{picture}(30,30)(0,0)
 \thicklines
\wavyline{-32}{25}{5} \put(-9.2,25){{\vector(1,0){0}}}
\put(-30,21){{\footnotesize Entangled}}
\put(-30,18){{\footnotesize Input State}}
\put(-9.5,20){\framebox(9,10){$U_{\theta}$}}
\put(-14,32){{{\footnotesize Q. Operation}}}
\wavyline{0}{25}{7} \put(30.8,25){{\vector(1,0){0}}}
\put(7,21){{\footnotesize Entangled}}
\put(7,18){{\footnotesize Output State}}
\put(26,32){{{\footnotesize Measurement}}} \put(30.5,10){\framebox(9.5,20){M}}
 \put(40.5,20){\vector(1,0){15}} \put(42,26){{\footnotesize Classical}} \put(42,23){{\footnotesize Data}}
\wavyline{-32}{15}{15}\put(30.8,15){{\vector(1,0){0}}}
\end{picture}
\end{center}
\caption{The use of a single entangled input state suffices.}
\label{fig:fig1}
\end{figure}

Another strategy is described in Refs. \cite{demartini:sqd,fujiwara:estsu2,acin:optestquantdyn}. It is enough to
use a single bipartite entangled state; one of the states is used as input for the quantum operation
and nothing is done to the other one, then the two qubits are measured,  as shown in Fig. \ref{fig:fig1}.
\medskip

In Ref.\cite{demartini:sqd} it is pointed out that in this setup there is a one to one correspondence between
the quantum operation and the joint output state. A maximally
entangled state is used as input and then the three components of the spin in both output particles are measured.
One can ask whether it is possible to
find a more accurate measurement. Also, is it possible to find a measurement that performs as well as the one in 
Ref. \cite{demartini:sqd}
which has less outcomes?
It will turn out that the answer is that one can find a more accurate measurement but this measurement is nonseparable. It
is also possible to find a measurement with less outcomes.

In Ref. \cite{fujiwara:estsu2} it is proven that a maximally entangled pure state is a
good input state in the sense that if $|\phi\ran \in \Cn^2 \otimes \Cn^2$ is maximally entangled, then
$$
\forall~\rho \textrm{ on } \Cn^2 \otimes \Cn^2,~H_\rho(\theta) \leq H_{|\phi\ran \lan \phi|}(\theta)
$$
where $H_\rho(\theta_0)$ is the quantum Fisher information matrix (QFI) for the joint output state, at $\theta=\theta_0$,
 if the input state is $\rho$. This quantity is defined for example in Ref. \cite{holevo:book} and explained in more detail in the following section.
  {The inverse of this matrix is a lower
bound (quantum Cram\'{e}r-Rao bound or QCRB) for the mean-square error of estimators based on arbitrary measurements of the output state}.
A maximally entangled state is a good input also in the sense
that the QCRB  can be achieved if and only if
the input state is maximally entangled. The problem here is that, as will be shown later, the measurement
that achieves the bound is actually a basis of projectors onto maximally entangled states.
This measurement can be performed using nonlinear optics but is far from being standard.

But perhaps the improvement in the estimation through the use of entangled measurements is not very large.
Are entangled measurements worth the trouble? The main aim of this paper is to show that the answer to this
question is positive. The value of entangled measurements will be quantified precisely.

Before continuing with this discussion it is necessary to explain what is meant
with ``better'' and ``best'' and how the quality of different
positive operator valued measures or POVMs is actually going to be quantified.

\section{\label{sec:defs} QFI, Fisher Information, and QCRB}
\subsection{Quantum Fisher information}
Suppose that the quantum state density matrix $\sigma$ on $\Cn^d$ is parametrized by
$\theta \in \Theta \subset \re^p$ where
$p$ is the number of parameters (less than or equal to $d^2-1$ for mixed states,
 $2d-2$ for pure states). In our case $\sigma$ would be the joint output state. Define the symmetric logarithmic
 derivatives $\lambda_1,\dots,\lambda_p$ as the
self-adjoint operators that satisfy
$$
\sigma_{,i}(\theta)=\partial_{\theta_i}\sigma(\theta)={\textstyle \half}[\sigma(\theta) \lambda_i(\theta) +\lambda_i(\theta) \sigma(\theta)].
$$
For pure states, $\sigma=|\psi\ran \lan \psi|$, they simply are $\lambda_i=2 \sigma_{,i}$.
The QFI is defined as the $p \times p$ matrix with elements
$$
H_{ij}(\theta)=\rlprt~\tr \left[ \sigma(\theta) \lambda_i(\theta) \lambda_j(\theta)\right]
$$
which for pure states reduces to
$$H_{ij}(\theta)=\rlprt \lan l_i(\theta) |l_j(\theta)\ran$$
where $|l_i(\theta)\ran=\lambda_i(\theta)|\psi(\theta)\ran$.
\subsection{(Classical) Fisher information}
Take a POVM with elements $M_1,\dots,M_n$. The Fisher information matrix (FI) for this measurement is
the $p \times p$ matrix with elements
$$
I_{ij}(M,\theta)=\sum_{\xi=1}^n \frac{\tr[\sigma_{,i}(\theta)M_{\xi}]\tr[\sigma_{,j}(\theta)M_{\xi}]}{\tr[\rho(\theta)M_{\xi}]}.
$$
For an estimator $\hat{\theta}$ and a measurement $M$, locally unbiased at $\theta_0$ \footnote{This means that
the expectation of the estimator satisfies $\esp_{M,\theta_0}(\hat{\theta}_i)={\theta_0}_i$
and $\eval{\partial_{\theta_j}\esp_{M,\theta}(\hat{\theta}_i)}{\theta =\theta_0} =\delta_{ij}$. },
the (classical) Cram\'er-Rao bound is satisfied
$$V(M,\theta_0,\hat{\theta}) \geq I(M,\theta_0)^{-1},$$
i.e., the FI is the smallest variance that a locally unbiased estimator based on this measurement can have. This
also means that if one of the eigenvalues of $I$ is zero, then the variance of the function
of the parameters corresponding to that eigenvalue is infinity and therefore cannot be
estimated.

If one has $N$ copies of the quantum state and performs the same measurement on each of the copies then the
FI of the $N$ copies, $I^N$, satisfies $I^N(M,\theta)=N I(M,\theta)$ where $I(M,\theta)$ is the FI of one system.
It follows that
$$V^N(M,\theta_0,\hat{\theta}) \geq {I^N}(M,\theta_0)^{-1}=I(M,\theta_0)^{-1}/N.$$
It is a well known fact in mathematical statistics that the maximum likelihood estimator (MLE) in the limit
of large $N$ is asymptotically unbiased and saturates the
classical Cram\'er-Rao bound. Moreover no other reasonable estimator
(unbiased or not) can do better.

\subsection{QCRB}
The QCRB states that for any measurement $M$
\be \label{eq:eqqcrb}
I(M,\theta) \leq H(\theta).
\ee
In other words, $H(\theta)-I(M,\theta)$ is a positive semidefinite matrix.

This bound is not achievable in general.
A theorem due to Matsumoto \cite{matsu:crb} states that for pure states,
the bound is achievable at $\theta=\theta_0$ if and only if
\be \label{eq:condqc}
\imprt \lan l_i(\theta_0) |l_j(\theta_0)\ran=0.
\ee
Furthermore, if condition (\ref{eq:condqc}) holds, there is a measurement with $p+2$
elements that achieves the bound.

In analogy with \cite{gillmassar:pra}, measurements will be compared using the quantity
$$\tr H(\theta)^{-1}I(M,\theta)$$
which is always less than or equal to $p$, the number of parameters.
For example, for the measurement used in \cite{demartini:sqd}, $\tr H^{-1}I=1$.

One needs to use a quantity like this because of the extra complexity
that quantum theory adds to the problem. Namely, in the most general case there is no
POVM that achieves equality in (\ref{eq:eqqcrb}). Typically, for any POVM $M_1$ which cannot be improved, one can find
another POVM $M_2$ such that neither $I(M_1,\theta) \leq I(M_2,\theta)$
nor $I(M_1,\theta) \geq I(M_2,\theta)$ are satisfied.
The bound (\ref{eq:eqqcrb}) is sharp, i.e., $H(\theta)$ is the smallest matrix larger than $I(M,\theta)$ for every $M$.
The difficulties can be overcome by using a
single number (instead of a matrix) to quantify the performance of a POVM. This defines an achievable bound
and any two POVMs can be compared according to this quantity. Of course, no single number
can be an \emph{absolute} quantification of the performance of a POVM. In applications
one must decide what one wants to estimate
and accordingly assign weights to the mean square error of the parameters to be estimated. This comes down to
using a quantity such as $\tr G(\theta) I(M,\theta)$. One needs to maximize this quantity for a general\footnote{This
quantity still has the property that if the inequality $\tr G(\theta) I(M_1,\theta) > \tr G(\theta) I(M_2,\theta)$
holds, then $I(M_1,\theta) \nless I(M_2,\theta)$.} $G(\theta) \geq 0$ tailored to one's specific needs. In this paper
the general problem is not solved. Only the case
$G(\theta)=H(\theta)^{-1}$ is considered. There are several good reasons for this choice:
\begin{enumerate}
\item  Since $H(\theta)$ is the smallest upper bound for all the $I(M,\theta)$, it defines a natural scale in which
to compare them.
 \item  $\tr H(\theta)^{-1}I(M,\theta)$ is parametrization invariant.
\item  $H(\theta)$ is closely related to the fidelity between true and estimated output states:
the metric generated by $H(\theta)$ is locally identical (up to a factor of 4) to the Bures distance,
$d_{\text{Bures}}(\rho,\sigma)^2=2(1-\sqrt{\mathcal{F}(\rho,\sigma)})$, where $\mathcal{F}$ is he fidelity,
which for pure states can be defined as $\mathcal{F}(|\psi\rangle \langle \psi|,|\phi\rangle \langle \phi|)=|\langle
\psi|\phi\rangle|^2$.
 \item Finally, the use of this quantity allows one to obtain simple and
striking results.
\end{enumerate}

\section{The case $d=2$}
\subsection{Entangled measurements}
In \cite{fujiwara:estsu2} it was shown that in this case the best input is any maximally entangled state. Here the
singlet state  $|\tau \ran =[|10\ran -|01 \ran]/\sqrt{2}$ will be used. The output is then
$|\psi(\alpha,\theta,\phi)\ran = [U(\alpha,\theta,\phi] \otimes \im) |\tau \ran$
where $U(\alpha,\theta,\phi)=\cos{\alpha}~\im+\iu \sin{\alpha}~\vec{n}_{\theta \phi}\cdot\vec{\sigma}$ is a $2 \times 2$
unitary matrix, $\vec{n}_{\theta \phi}$ is the unit vector $(\sin{\theta}\cos{\phi},\sin{\theta}\sin{\phi},\cos{\theta})$
parametrized by its polar coordinates, and $\vec{\sigma}=(\sigma_1,\sigma_2,\sigma_3)$ are the Pauli matrices.

It is quite straightforward to calculate that the QFI is
$$
H(\alpha,\theta,\phi)= 4\left (\begin{array}{ccc}
1 & 0   &0 \\
0 &(\sin \alpha)^2 & 0 \\
0&0&(\sin \alpha \sin \theta)^2
\end{array} \right )
$$
and as expected $\imprt \lan l_i(\alpha,\theta,\phi) |l_j(\alpha,\theta,\phi)\ran=0$. One can find a simple measurement
that achieves the bound; in fact, any measurement of the type
\be
&M_{\alpha}=|b_{\alpha}\ran \lan b_{\alpha}|~~\alpha=1,\dots, p+1, \\
&M_{p+2}=\im-\sum_{\alpha=1}^{m+1} M_{\alpha}, \\
&|b_{\alpha}\ran=\sum_{\beta=1}^{p+1} o_{\alpha \beta}|m_{\beta}\ran,\\
&|m_k\ran=\sum_l (H^{-1/2})_{kl} |l_l\ran,~~
|m_{p+1}\ran=|\phi\ran,
\ee
with $o$ a $(p+1)\times(p+1)$ real orthogonal matrix satisfying $o_{\alpha,p+1}\neq 0$
achieves the bound. For example measuring the Bell basis
\be \label{eq:bellbasis}
M_1^{Bell}&= \frac{|00\ran - |11 \ran}{\sqrt{2}}\frac{\lan 00| - \lan 11 |}{\sqrt{2}},& \\
M_2^{Bell}&= \frac{|00\ran + |11 \ran}{\sqrt{2}}\frac{\lan 00| + \lan 11 |}{\sqrt{2}},& \\
M_3^{Bell}&= \frac{|01\ran + |10 \ran}{\sqrt{2}}\frac{\lan 01| + \lan 10 |}{\sqrt{2}},& \\
M_4^{Bell}&= \frac{|01\ran - |10 \ran}{\sqrt{2}}\frac{\lan 01| - \lan 10 |}{\sqrt{2}}&
\ee
achieves $I(M^{Bell},\theta)=H(\theta)$ for all $\theta$ and therefore satisfies
\be
\tr H^{-1}(\theta)I(M^{Bell},\theta)=3
\ee
everywhere; this is three times the value achieved in \cite{demartini:sqd}.
This measurement, which has been implemented using non-linear optics \cite{Shih:CompBellMeas},
is not widely available in quantum optics labs.
On the other hand, a POVM with the two components
\be \label{eq:redbellbasis}
M_1&= M_k^{Bell},& \\
M_2&=\im -M_k^{Bell} &
\ee
for $k=1,2,3$ or $4$ has been implemented with linear optics
and is much more standard than measuring the whole basis. This is a POVM with only two outcomes
(which might be an advantage for its practical implementation) and calculation
shows that it satisfies $\tr H^{-1}I=1$
everywhere, which is as good as the measurement in \cite{demartini:sqd}. It
will be shown in the next section that it is actually as good as any separable measurement (in terms of
the value of $\tr H^{-1}I$).
This measurement does have a serious drawback, namely that one can only identify one function of
$\alpha$, $\beta$ and $\phi$.  {The drawback can be overcome, for example,
by measuring (\ref{eq:redbellbasis}) for $k=1,2$ and $3$ each in one third of the available copies. In this way one is
 able to identify all three parameters, $\tr H^{-1}I=1$ is still satisfied,
and finally this new POVM should not be harder to implement than the previous one. Again here everything
depends on what one wants to estimate.

A measurement with three elements
\be
M_1&= M_k^{Bell},& \\
M_2&=M_l^{Bell}, & \\
M_3&=\im - M_k^{Bell} - M_l^{Bell} &
\ee
for some $k \neq l$ has also been implemented with linear optics. In fact, it has been shown
\cite{Calsamiglia:MaxEffBellMeas} that, with linear optics, this is the best one can do. This POVM
satisfies $\tr H^{-1}I=2$. This is twice the value that can be achieved
with any separable measurement. Note that this measurement has the same weakness as the previous one: it cannot identify
all three parameters (it identifies two functions of $\alpha$, $\theta$ and $\phi$). It is easily possible to overcome
this difficulty in a similar way as before.}
\subsection{LOCC measurements}

{How well can one estimate $U$ using only LOCC meas\-ure\-ments:
measurements that can be implemented locally
and with the aid of classical communication between the two parties.
In fact, to begin with, the larger class of \emph{separable} measurements will be studied: measurements
whose elements are positive combinations of products of one dimensional projectors. This definition of ``separable''
is a slight generalization of that of \cite{bennet:NonLocWithoutEnt}, where it was shown that
 there exist separable measurements which are not LOCC. Nonseparable measurements are called entangled.}

Consider a separable POVM with elements
$M_\xi= \sum_i c_{\xi i} (|\psi_{\xi
i}^A\ran \otimes |\psi_{\xi i}^B\ran) (\lan \psi_{\xi i}^A|\otimes
\lan \psi_{\xi i}^B|)$. One can \emph{refine} this POVM to
obtain another POVM with elements that are proportional to
one-dimensional projectors $M_{\xi i}= c_{\xi i} (|\psi_{\xi i}^A\ran \otimes
|\psi_{\xi i}^B\ran) (\lan \psi_{\xi i}^A|\otimes \lan \psi_{\xi
i}^B|)$. By relabeling $\xi~i \to \xi$ one obtains
\be \label{eq:sepmeas}
M_{\xi}= c_{\xi}(|\psi_{\xi}^A\ran \otimes
|\psi_{\xi}^B\ran) (\lan \psi_{\xi}^A|\otimes \lan
\psi_{\xi}^B|).
\ee
The Fisher information corresponding to this
refined POVM is greater than or equal to the Fisher information of
the original POVM. Thus, since one wants to maximize the FI, one may
restrict oneself to measurements of the type described in
Eq. (\ref{eq:sepmeas}).

For the calculations that follow it is more convenient to express Eq. (\ref{eq:sepmeas}) using the
Pauli matrices
$$M_\xi = c_\xi \frac{\im +\vec{a}_\xi\cdot \vec{\sigma}}{2}\otimes\frac{\im +\vec{b}_\xi\cdot \vec{\sigma}}{2}$$
with $|\vec{a}_\xi|=|\vec{b}_\xi|=1$ and $c_\xi>0$.
The condition $\sum_\xi M_\xi=\im$ can be rewritten as follows:
\be \label{eq:conds}
&\sum_{\xi}c_\xi =4,
\sum_{\xi}c_\xi \vec{a}_\xi=0, \\
&\sum_{\xi}c_\xi \vec{b}_\xi=0 ,
\sum_{\xi}c_\xi a_{\xi k}b_{\xi l}=0.
\ee
Since
$$ \frac{|10\ran -|01\ran}{\sqrt{2}}\frac{\lan10| -\lan01|}{\sqrt{2}}
=\frac{1}{4}\left(\im \otimes \im -\sum_{i=1}^3 \sigma_i \otimes \sigma_i \right) $$
the density matrix of the system can be written as
$$\rho(\alpha,\theta,\phi)=\frac{1}{4}\left(\im \otimes \im -\sum_{i=1}^3
U(\alpha,\theta,\phi)\sigma_i U^{\dag}(\alpha,\theta,\phi) \otimes \sigma_i \right).$$
The probabilities are then
$$p_\xi=\frac{c_\xi}{4}\left(1 -\sum_{i,j=1}^3 b_{\xi i} a_{\xi j}\frac{\tr{(U\sigma_iU^{\dag}\sigma_j})}{2}  \right)$$
and $\half\tr{(U\sigma_iU^{\dag}\sigma_j)}$ can be calculated to be
$$
\cos{2\alpha} ~\delta_{ij}- \sin 2 \alpha \sum_{k=1}^3 \epsilon_{ijk}n_k + 2 \sin^2{\alpha}~n_i n_j.
$$
Substituting this into the expression for the probabilities one gets
\be
& p_\xi=\frac{c_\xi}{4} \left(1-\cos{2\alpha}~(\vec{a}_\xi \cdot \vec{b}_\xi)+
\sin{2\alpha}~(\vec{b}_\xi \times \vec{a}_\xi \cdot \vec{n}) \right. \\ & \left. -
2 \sin^2{\alpha}~(\vec{n} \cdot \vec{a}_\xi)(\vec{n} \cdot \vec{b}_\xi ) \right)
.\ee
After some not very interesting manipulations one finds
\be
&\frac{1}{p_{\xi}}\left( \frac{(p_{\xi,\alpha})^2}{4}+\frac{(p_{\xi,\theta})^2}{4 \sin^2{\alpha}}+
\frac{(p_{\xi,\phi})^2}{4 \sin^2{\alpha}~\sin^2{\theta}} \right) \\
&=\frac{c_\xi}{4} \left(1+\cos{2\alpha}~(\vec{a}_\xi \cdot
\vec{b}_\xi)- \sin{2\alpha}~(\vec{b}_\xi \times \vec{a}_\xi \cdot
\vec{n})  \right. \\ & \left. +
 2 \sin^2{\alpha}~(\vec{n} \cdot \vec{a}_\xi)(\vec{n} \cdot \vec{b}_\xi ) \right)
.
\ee
Finally, using the conditions (\ref{eq:conds}) one obtains that for separable measurements of the type (\ref{eq:sepmeas})
\be
&\tr{   [H^{-1}(\theta)I(M,\theta)]} \\
&=\sum_\xi \frac{1}{p_{\xi}}\left( \frac{(p_{\xi,\alpha})^2}{4}+\frac{(p_{\xi,\theta})^2}{4 \sin^2{\alpha}}+
\frac{(p_{\xi,\phi})^2}{4 \sin^2{\alpha}~\sin^2{\theta}} \right) \\ &=1.
\ee

{Any separable measurement can be refined to a measurement of the type (\ref{eq:sepmeas}).
Therefore for all separable measurements $M_{\textrm{sep}}$ $$\tr H(\theta)^{-1}
I(M_{\textrm{sep}},\theta) \leq 1.$$
This bound therefore also holds for LOCC measurements. Since there are LOCC measurements of the type
(\ref{eq:sepmeas}), the bound is achievable with LOCC measurements.}

\section{The case $d>2$}
\subsection{Entangled Measurements}
Before starting with any calculations it will be shown that the
quantity that is being analyzed
$$f(\theta)=\sup_{M} \tr H^{-1}(\theta) I(\theta,M)$$ does not depend on $\theta$.

For any $\theta_1$ and $\theta_0$ there exists a unitary matrix $V$ such that $VU(\theta_0)=U(\theta_1)$.
It is easy to see that for such a choice
$$
\tr H^{-1}(\theta_1) I(\theta_1,(V \otimes \im)M(V \otimes \im)^{\dag})=\tr H^{-1}(\theta_0) I(\theta_0,M)
$$
This implies that
$$
\sup_{M_1}\tr H^{-1}(\theta_1) I(\theta_1,M_1) \geq \sup_{M_0}\tr H^{-1}(\theta_0) I(\theta_0,M_0).
$$
Thus $f(\theta_1)\geq f(\theta_0)$, but since $\theta_0$ and $\theta_1$ are arbitrary, the function
$f$ must be constant. Therefore one can choose any value of the parameter to perform the
calculations. One of the implications this has is that if one proves that the QCRB can (not) be achieved
at one value of the parameter, then it can (not) be achieved everywhere (anywhere).

In \cite{fujiwara:estsu2} it is mentioned that in dimension $d>2$, it is no longer true
 that a maximally entangled state maximizes the QFI;
however it is still true that the QCRB is achieved if and only if
the input state is maximally entangled. In order to prove the first
statement it is enough to find a counter example. This is
not difficult to do for example in $d=3$. The second statement is also not
difficult to prove and because of the
last discussion it will be enough to do it for $U$ equal to the identity.

A $SU(d)$ matrix can be written as
$\exp \left( \iu \sum_{\alpha=1}^{d^2-1} \theta_\alpha T_\alpha \right )$.
Here $\theta \in \re^{d^2-1}$ and the $T$'s are
in the $su(d)$ Lie Algebra.
They are traceless self-adjoint matrices and are chosen so that they also satisfy:
$$\tr(T_\alpha T_\beta)=\delta_{\alpha \beta}.$$
For $U$ close to the identity (or $\theta$ close to zero),
$$U \approx \im +\iu \sum_{\alpha=1}^{d^2-1} \theta_\alpha T_\alpha$$
the input state can be written as
$\sum_{kl} R_{kl}|kl\ran$.
Normalization implies $\tr RR^{\dag}=1$ where
$R$ is the $d \times d$ matrix with elements $R_{kl}$. Since
$RR^\dag$ has trace one and is self-adjoint it can be written
$RR^\dag=\im /d+\sum_{\alpha} t_\alpha T_\alpha$
where the $t$'s are real numbers.
At the identity the output state satisfies
 \be
 |\psi\ran & = \sum_{kl} R_{kl}|kl\ran, \\
 |\psi_{,\alpha}\ran &= i \sum_{kl} R_{kl} T_{\alpha}|k\ran \otimes |l\ran;
 \ee
the $|l_\alpha \ran$ vectors defined in section \ref{sec:defs} can be written as
$$|l_\alpha\ran=2(|\psi_{,\alpha}\ran+\lan \psi_{,\alpha}|\psi\ran |\psi \ran)$$
 and the condition for achieving the QCRB (\ref{eq:condqc}) becomes
\be \label{eq:condqcsud}
\imprt \lan l_\alpha |l_\beta \ran=4\imprt \lan \psi_{,\alpha}|\psi_{,\beta} \ran
=\frac{2 \tr (RR^{\dag}[T_\alpha, T_\beta])}{\iu} =0
\ee
for all $\alpha$ and $\beta$. This implies $RR^\dag=\im /d$ \footnote{Since $(RR^\dag-\im /d) \in su(n)$
and $su(n)$ is a perfect Lie algebra (i.e. can be spanned by commutators), Eq. (\ref{eq:condqcsud}) may be rewritten as
$\forall_{Y \in su(n)}\tr [(RR^\dag-\im /d)Y ]=0,$
this implies $RR^\dag-\frac{\im}{d}=0$ because the trace form is non-degenerate.}. This means that the input
 state is maximally entangled \footnote{The condition for a bipartite state to be maximally entangled
 is that the partial trace should be proportional to the identity. In our case $\tr_2 |\psi \ran \lan \psi|=RR^\dag$.}.

For the calculations the maximally entangled state, $\sum_{k=1}^d|kk\ran/ \sqrt{d}$, is used.
$H$ can be very easily calculated to be
\be
H_{\alpha \beta}&=\frac{4}{d} \delta_{\alpha \beta}.
\ee
Since the QCRB can be achieved
$$\sup_M \tr H^{-1}(\theta)I(M,\theta) = d^2-1.$$

\subsection{LOCC measurements}

It will be shown here that for all separable measurements $M_{\textrm{sep}}$ the following holds
\be \label{eq:loccbound}
\tr H^{-1}(\theta)I(M_{\textrm{sep}},\theta) \leq \frac{d(d-1)}{2}. \ee
This shows that if one allows nonseparable measurements, the estimation can be improved
by a factor of $2(d+1)/d$ with respect to separable measurements. This is
always more than twice.

In order to prove Eq.(\ref{eq:loccbound}) a particular representation for the
$T$'s will be chosen, namely:
\be
&T_{kls}=\iu^s \frac{|k\ran \lan l|+(-1)^s |l\ran \lan k|}{\sqrt{2}}&~~k>l,~s=\{0,1\},\\
&T_m= \sum_{k=1}^d c_{mk} |k\ran \lan k|&~~m=1,\dots, d-1,
\ee
where the coefficients $c_{mk}$ obey
\be
&\sum_{k=1}^d c_{mk}=0, \\
&\sum_{k=1}^d c_{mk} c_{nk}= \delta_{mn}.
\ee
From these two one can derive the relation
\be
&\sum_{m=1}^{d-1} c_{mk} c_{ml}= \delta_{kl}-\frac{1}{d}.
\ee
Measurements of the form
$$M_\xi=c_\xi |\phi_\xi \ran \lan \phi_\xi | =c_\xi |a_\xi \ran \lan a_\xi | \otimes |b_\xi \ran \lan b_\xi |$$
are considered. The quantity of interest is
\be \label{eq:trace}
& \tr H^{-1}I=\frac{d}{4}\tr I  \\
& =\frac{d}{4}\sum_{\xi \alpha} c_\xi \frac{(\lan \phi_\xi|\psi_{,\alpha}\ran \lan \psi|\phi_\xi\ran
+\lan \phi_\xi|\psi\ran \lan \psi_{,\alpha}|\phi_\xi\ran)^2}
{|\lan \phi_\xi|\psi\ran|^2}\\
&=\frac{d}{2} \sum_\xi c_\xi \left[ \rlprt \left( \frac{\lan \psi|\phi_\xi\ran}{\lan \phi_\xi|\psi\ran}\sum_{\alpha=1}^{d^2-1}
\lan \phi_\xi|\psi_{,\alpha}\ran^2 \right)  \right. \\ & \left. +
 \sum_{\alpha=1}^{d^2-1}
\lan \phi_\xi|\psi_{,\alpha}\ran \lan \psi_{,\alpha}|\phi_\xi\ran   \right].
\ee
The second term in the previous equation is easy to calculate,
\be \label{eq:secondterm}
&\frac{d}{2} \sum_\xi c_\xi   \sum_{\alpha=1}^{d^2-1}
\lan \phi_\xi|\psi_{,\alpha}\ran \lan \psi_{,\alpha}|\phi_\xi\ran
=\frac{d}{2}  \sum_{\alpha=1}^{d^2-1}
 \lan \psi_{,\alpha}|\psi_{,\alpha}\ran  \\  &= \frac{d}{2}  \sum_{\alpha=1}^{d^2-1}
 \frac{\tr T_\alpha^2}{d} =\frac{d^2-1}{2},
\ee
but for the first term a little more work will be needed. One needs to calculate
$$\lan \phi_\xi|\psi_{,\alpha}\ran=\frac{i}{\sqrt{d}}\sum_{k=1}^d\lan a_\xi|T_\alpha|k\ran \lan b_\xi|k\ran.$$
For $\alpha=\{kls\}$
\be
&\lan \phi_\xi|\psi_{,kls}\ran =\frac{i^{s+1}}{\sqrt{2 d}} [\lan a_\xi|k\ran \lan b_\xi|l\ran +(-1)^s
\lan a_\xi|l\ran \lan b_\xi|k\ran], \\
&\sum_{s=0}^1\lan \phi_\xi|\psi_{,kls}\ran^2=-\frac{2}{d}\lan a_\xi|k\ran \lan b_\xi|l\ran \lan a_\xi|l\ran \lan b_\xi|k\ran
.\ee
Since the last expression is symmetric with respect to exchanging $k$ with $l$,
$\sum_{k>l}=\half \sum_{k\neq l}=\half(\sum_{kl}-\sum_{k=l})$ and
$$
\sum_{k>l} \sum_{s=0}^1\lan \phi_\xi|\psi_{,kls}\ran^2=\frac{1}{d}\sum_{k=1}^d \lan a_\xi|k\ran^2 \lan b_\xi|k\ran^2
-\lan \phi_\xi|\psi\ran^2.
$$

In the case $\alpha=m$
\begin{align*}
&\sum_{m=1}^{d-1}\lan \phi_\xi|\psi_{,m}\ran^2\\
&=-\frac{1}{d} \sum_{k,l=1}^d \sum_{m=1}^{d-1} c_{mk} c_{ml}
 \lan a_\xi|k\ran \lan b_\xi|k\ran  \lan a_\xi|l\ran \lan b_\xi|l\ran \\
&= \frac{1}{d} \sum_{kl} (\frac{1}{d}-\delta_{kl}) \lan a_\xi|k\ran \lan b_\xi|k\ran \lan a_\xi|l\ran \lan b_\xi|l\ran\\
&=\frac{1}{d}\lan \phi_\xi|\psi\ran^2-\frac{1}{d}\sum_{k=1}^d \lan a_\xi|k\ran^2 \lan b_\xi|k\ran^2
\end{align*}
putting things together
\be
\sum_{\alpha=1}^{d^2-1}\lan \phi_\xi|\psi_{,\alpha}\ran^2&=\frac{1-d}{d}\lan \phi_\xi|\psi\ran^2
\ee
and
\be
&\frac{d}{2} \sum_\xi c_\xi \rlprt \left( \frac{\lan \psi|\phi_\xi\ran}{\lan \phi_\xi|\psi\ran}\sum_{\alpha=1}^{d^2-1}
\lan \phi_\xi|\psi_{,\alpha}\ran^2 \right)\\
&=\frac{1-d}{2}  \sum_\xi c_\xi |\lan \psi|\phi_\xi\ran|^2=\frac{1-d}{2}
.\ee
Finally, substituting  the previous equation and Eq. (\ref{eq:secondterm}) in (\ref{eq:trace}) one obtains
the desired result, namely, for any separable measurement $M$ of the type (\ref{eq:sepmeas})
\be
\tr H^{-1}(\theta)I(M,\theta)=\frac{d(d-1)}{2}
.\ee
Of course this impliesEq. (\ref{eq:loccbound}). The argument for LOCC measurements is the same as for the two
dimensional case and one obtains the same bound for them.
\section{Conclusions and open problems}
In this paper it has been shown that by using nonseparable measurements
there is a significant improvement in the accuracy of the estimation of unitary operations.
It is also proven that in $d$ dimensions the QCRB can be achieved
if and only if the input state is maximally entangled. An open problem is
the estimation of more general quantum operations, described by the \emph{Kraus decomposition} \cite{Nielsen:book}.
\begin{acknowledgments}
This research was funded by the Netherlands Organization for
Scientific Research (NWO), support from the RESQ (IST-2001-37559)
project of the IST-FET programme of the European Union is also
acknowledged.
\end{acknowledgments}

\end{document}